\documentstyle[twocolumn,epsf,prl,aps]{revtex}
\newcommand {\be}{\begin{equation}}
\newcommand {\ee}{\end{equation}}
\newcommand {\bea}{\begin{eqnarray}}
\newcommand {\eea}{\end{eqnarray}}
\begin{document}
\draft

\title
{Quantum Creep in $\bf{Y_{1-x}Pr_{x}Ba_{2}Cu_{3}O_{7-\delta}}$ Crystals: Magnetic Relaxation and
Transport.}

\author{T. Stein, G. A. Levin,  and C. C. Almasan}
\address{Department of Physics, Kent State University, Kent  OH 44242}
\author{D. A. Gajewski and M. B. Maple}
\address{Department of Physics and Institute for Pure and Applied Physical
Sciences, University of California, San Diego, La Jolla, CA 92093
%\date{\today}
\vspace{.25cm}
%\maketitle
{\rm \begin{quote}
%\begin{abstract}
\hspace{.25cm} We report transport and magnetic relaxation measurements in the mixed state of
strongly underdoped $Y_{1-x}Pr_{x}Ba_{2}Cu_{3}O_{7-\delta}$  crystals.  A transition from thermally
activated flux creep to temperature independent quantum flux creep is observed in both transport and
magnetic relaxation at temperatures $T\le  5\;K$.  Flux transformer measurements indicate that the
crossover to quantum creep is preceded by a coupling transition.  Based on these observations we
argue that below  the coupling transition the current is confined within a very narrow layer
beneath the current contacts.
%\end{abstract}
\end{quote}
}
}
\maketitle
%\pacs{Pacs: 74.60.Ge, 74.50.+r, 74.72.Bk}

\narrowtext
Quantum flux creep in superconductors is the only experimentally accessible phenomenon in  which a
macroscopic metastable   system, such as a persistent supercurrent, relaxes coherently without
thermal activation. In all other  metastable systems, the relaxation proceeds as a sequence of a
large number of uncorrelated microscopic steps, requiring thermal activation over an energy
barrier. Yet, the evidence of magnetic  relaxation and resistance  that do not extrapolate to zero
at a temperature $T\rightarrow 0$ has  not reached a point where experimental data  begin to form a
cohesive picture of the phenomenon. While the non-vanishing magnetic relaxation has  been  observed
both in single crystals and thin films \cite{Mota1,vanDalen,Hoekstra}, the non-vanishing temperature
independent resistance  had been observed only in ultrathin films and multilayers
\cite{Y.Liu,Ephron,Chervenak}. This has contributed to the assertion that the temperature independent resistance in films and the non-vanishing low-temperature magnetic
relaxation in single crystals are unrelated phenomena. 

A second question, closely related to quantum creep in high-$T_c$ superconductors, is the coupling
transition in layered crystals.  Several groups have arrived at conflicting conclusions.  Safar et
al.\cite {H.Safar} reported that the voltages generated by the motion of vortices on the opposite
faces of the $YBa_2Cu_3O_{7-\delta}$ sample (flux transformer method) converge.  However, such a
strong manifestation of the  coupling transition appears to be the exception rather than the rule.
Other groups have observed that the voltages generated on the opposite faces of  $Bi_2Sr_2CaCu_2O_8$
\cite{R.Busch,Safar2,Doyle,Keener} and $YBa_2Cu_3O_{7-\delta}$ \cite{Y.Eltsev} crystals diverge
with lowering temperature, rather than converge.

Here we report the results of both magnetic relaxation and transport measurements on strongly
underdoped crystals of $Y_{0.47}Pr_{0.53}Ba_{2}Cu_{3}O_{7-\delta}$ ($T_c\approx 17-21\;K$).  Our
transport data indicate that the onset of quantum creep is preceded by a a coupling transition 
which leads to non-ohmic dissipation. The transition to temperature independent creep takes place
at lower $T$ and  only in a very thin layer below the current contacts;  the rest of the  
\hspace{-0.15in} sample continues to exhibit thermally activated creep.  A picture of the low
temperature dissipation of the transport current that arises from these observations is that the
sample is divided into two macroscopic regions: a very thin layer, a few unit cells thick, near the
upper face, where the current contacts are located, which carries most of the current, and the rest
of the  sample which remains mostly undisturbed by the current.  These two macroscopic regions are
decoupled from each other. Inside these layers the vortices are coherent, with the correlation
length comparable to the thickness of the respective layer. Magnetic relaxation measurements
confirm the transition to T-independent relaxation at approximately the same temperatures  as in 
transport.      

\begin{figure}
%\resizebox{\columnwidth}{!}{\includegraphics*[.5in,5.9in][4in,10.8in]{fluxlet_FIG_1.eps}}
%\hspace*{-.2in}
%\special{illustration fluxlett_fig_1.eps}
\epsfxsize=\columnwidth \epsfbox{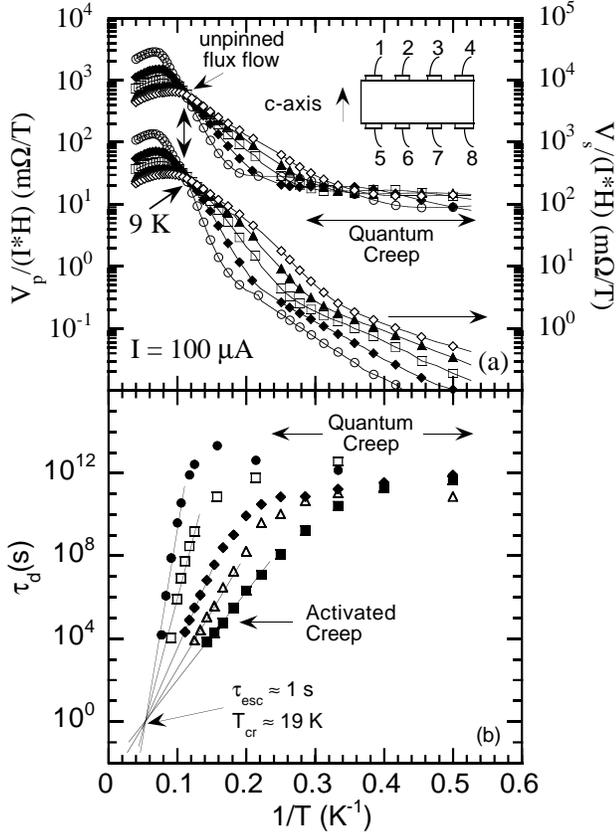}
%\vspace*{-6in}
\caption{{\bf (a)} Primary $V_{p}$ and secondary $V_{s}$ voltages, normalized to the
total current $I$ and magnetic field $H$, plotted vs $1/T$ for five magnetic fields ($0.2, 0.4,
0.6, 0.8$ and $1\;T$).  The slope decreases with increasing field. {\bf Inset:}  Contact
configuration used in the measurements.  {\bf (b)} Decay time $\tau_{d}$ of magnetic moment (see
definition in text)  vs $1/T$ for several values of magnetic field ($H=0.1$, $0.2$, $0.6$, $0.8$,
and $1.2\;T$). The slope decreases with increasing field. The straight line extrapolations of the 
Arrhenius-type dependence  converge at $T_{cr}\approx 18-19\;K$ and $\tau_{esc}\approx 1-2\;s$. The
saturation of $\tau_d$ at the level $10^{11}-10^{12}s$ is due to quantum creep.}
\label{alldata}
\end{figure}

Two strongly underdoped twinned single crystals of $Y_{0.47}Pr_{0.53}Ba_{2}Cu_{3}O_{7-\delta}$ with
$T_c\approx 17$ and $21\;K$, respectively, were prepared as described in Ref.~\cite{L.M.Paulius}.
The first sample was used for transport measurements with  the ``flux transformer'' contact
configuration  [inset to Fig.~\ref{alldata}(a)]: The current $I$ was injected through the contacts
on one face of the sample and   the voltage drop between  the voltage contacts on the same (primary
voltage $V_p$) and the opposite (secondary voltage $V_s$) faces was measured for temperature, total
current, and magnetic field $H$, applied parallel to the c-axis, in the ranges
$1.9\;K\le T\le25\;K$, $0.3\;\mu A \le I\le 2\;mA$, and $0.2\;T\leq H\leq 9\;T$.  Magnetic
relaxation measurements were performed on the second crystal using a  $SQUID$ magnetometer.
The crystal was cooled in zero field; a field $H+\Delta H$ ($\Delta H=0.3\;T$ for all $H$) was
applied parallel to its c-axis, and then the field was reduced to $H$. The decay of the resultant
paramagnetic moment was monitored  for several hours ($\approx 10^4\;s$) in constant $H$.

Figure~\ref{alldata}(a) gives an overview of the $T$ and $H$ dependence of $V_p$ and $V_s$
normalized to $I$ and $H$.  The convergence of these curves at $T \approx 9\;K$ indicates a regime
of free flow of vortices,  where $V_{p,s}/I\propto H$.  In a region below $9\;K$, $V_{p}$ and
$V_{s}$ exhibit activated $T-$dependences with field-dependent activation energies.  At even lower
%\vspace*{10.1in}
%
\hspace*{-.125in}$T$, the primary signal becomes T-independent and scales  approximately with $H$,
while the secondary voltage remains thermally activated.

\begin{figure}
%\resizebox{\columnwidth}{!}{\includegraphics*[.5in,5.85in][3.6in,10.8in]{fluxlett_FIG_2.eps}}
%\hspace*{-.2in}
\epsfxsize=\columnwidth \epsfbox{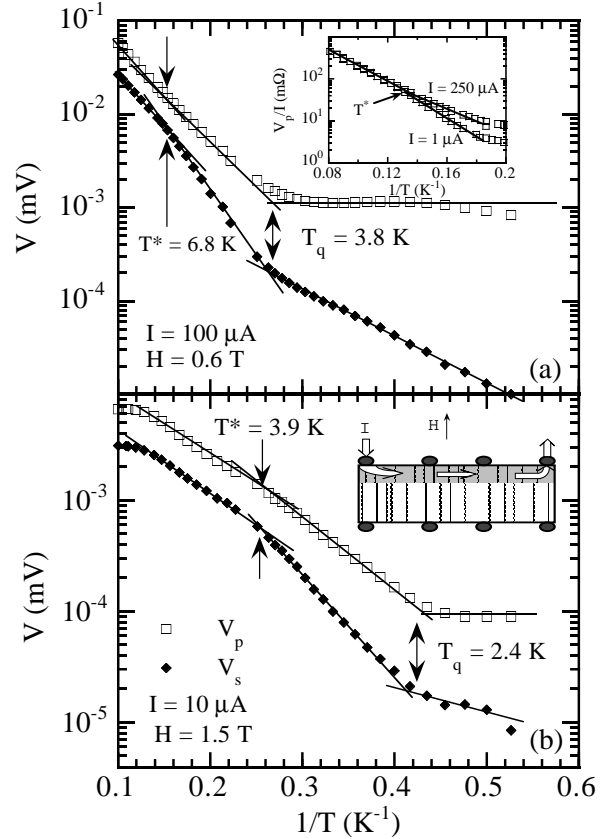}
%\special{illustration fluxlett_fig_2.eps}
%\vspace*{-6in}
\caption{Arrhenius plots of the primary $V_{p}$ and secondary $V_s$ voltages measured in two 
different fields and currents: 
{\bf (a)} $H=0.6\;T$ and $I=100\; \mu A$, and 
{\bf (b)} $H=1.5\;T$ and $I=10\; \mu A$. 
{\bf Inset to (a):} Expanded view of the primary resistance $V_p/I$ for two  values of the total
current ($1\mu A$ and $250\mu A$)  in an applied magnetic field $H=0.2\;T$.
{\bf Inset to (b):} Schematic representation of the current distribution. The current
flows mainly in the upper (shaded) layer. The vortices are coupled within each layer,
but the layers are decoupled from each other.}
\label{Arhenn}
\end{figure}

Figures~\ref{Arhenn}(a) and \ref{Arhenn}(b) represent an expanded view of $V_{p,s}$ vs $1/T$ for
$H=0.6\;T$, $I=100\;\mu A$ and $H=1.5\;T$, $I=10\; \mu A$, respectively. The activation energies
$E_{p,s}\equiv-d\ln V_{p,s}/d(1/T)$ of both $V_p$ and $V_s$ change suddenly at a temperature
$T^*(H)$. In addition, the dissipation becomes  strongly  non-ohmic below $T^*$. As the inset to
Fig.~\ref{Arhenn}(a) shows, the primary resistance $R_p(T)\equiv V_{p}/I$ is current independent
only above $T^*$.  The activation energies $E_{p,s}$, which are equal and current-independent at
$T>T^*$, become current-dependent at $T<T^*$. At low current, the activation energy  below $T^*$ is 
{\it greater } than above $T^*$ [Fig.~\ref{Arhenn}(b)] so that the curves $V_{p,s}(1/T)$ have a
downward curvature.  However, the activation energy decreases with increasing current and $E_{p}(I)$
becomes smaller than it is
%\vspace*{10.1in}
%
\hspace*{-.125in}at $T>T^*$, so that  $V_{p}(1/T)$ acquires an upward curvature [$V_p$
in Fig.~\ref{Arhenn}(a) and the inset].
 
This evidence indicates that the vortices undergo a coupling transition at $T^*$. At $T>T^*$, the
dissipation mechanism is activated hopping of 2D pancakes over potential barriers since the
activation energies are the same for the primary and secondary voltages, in spite of the nonuniform
current distribution. When the pancakes begin to form coherent lines, the activation energy
increases so that $V_{p,s}(1/T)$ curve downward. However, the fact that $V_p$ remains greater than
$V_s$ (below $T^*$, $V_p/V_s$ rapidly increases with decreasing T) indicates that the vortices do
not extend through the whole sample. 

\begin{figure}
%\resizebox{\columnwidth}{!}{\includegraphics*[.5in,8.1in][3.6in,10.8in]{fluxlett_FIG_3.eps}}
%\hspace*{-.2in}
%\special{illustration fluxlett_fig_3.eps}
\epsfxsize=\columnwidth \epsfbox{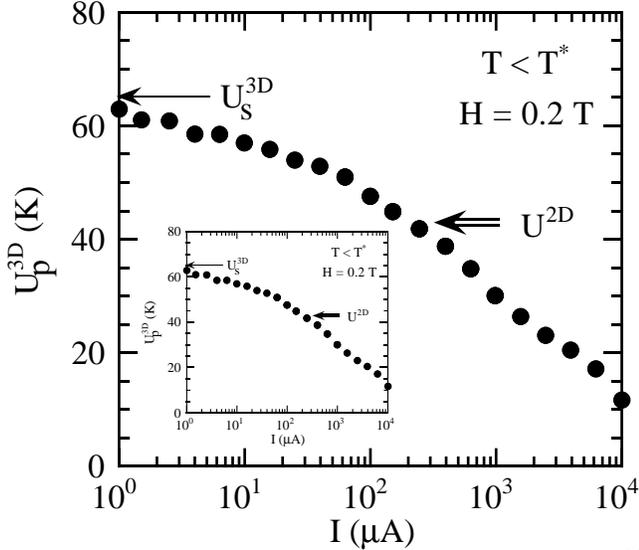}
\vspace*{-1.5in}
\caption{Current dependence of the activation energy $U^{3D}_p$ determined  as $d\ln V_{p}/d\ln (1/T)$ for
$T<T^*$ in a field $H=0.2\;T$. In the limit of small current, $U^{3D}_p\approx U^{3D}_s$ as shown
by the arrow. The value of $U^{2D}\approx 45\;K$ is also indicated by the double arrow.
{\bf Inset:} Field dependence of the activation energy $U^{3D}_s$ determined as 
$d\ln V_{s}/d\ln (1/T)$ for $T<T^*$.  The solid line corresponds to $H^{-1/2}$ dependence.}
\label{fig:U(I)}
\end{figure}

An important fact for understanding the nature of dissipation below  $T^*$ is that the activation
energy decreases with increasing current and eventually becomes {\it smaller} than that
for 2D pancakes.  Figure~\ref{fig:U(I)} displays the current dependence of $U^{3D}_p\equiv E_{p}$ at
$T<T^*$ for $H=0.2\;T$.  The
%\vspace*{10.05in}
%
\hspace*{-0.125in}double arrow indicates the value of the current independent
$U^{2D}\equiv E_{p,s}$ at $T>T^*$ and for the same magnetic field. The fact that $U^{3D}_p<U^{2D}$ for $I>0.2\;mA$
indicates that the vortices remain coupled even at large  currents which allows a new channel of
relaxation to emerge and become dominant due to smaller activation energy. The inset to
Fig.~\ref{fig:U(I)} is a log-log plot of $U^{3D}_s$ vs $H$, obtained from the secondary voltage
$V_{s}$ (limit of  very small current). The values of $U^{3D}_s$ follow an $H^{-1/2}$ dependence. This field dependence  of
$U^{3D}_s$ is consistent with a $3D$ plastic creep model, where $U^{3D}_s$ is estimated as the energy needed for the formation of a double kink over the Peierls barrier\cite{G.Blatter:Review,Y.Abulafia}.  These results show that
the dissipation at $T<T^{*}$ is determined by two parallel processes: thermally activated motion of
correlated vortices (dominant at low currents) with the activation energy greater than that for 2D
pancake vortices, and plastic motion of dislocations (dominant at higher currents) with the
activation energy smaller than that for a 2D vortex.

At lower  temperatures, the primary voltage $V_{p}$ becomes temperature independent for $T<T_q(H)$
and scales approximately with the field [Fig.~\ref{alldata}(a)] while $V_s$ remains thermally
activated, but with a noticeably smaller activation energy (Fig.~\ref{Arhenn}). 

The high normal state resistivity of this specimen may be the reason that quantum creep begins  to
dominate the thermally-activated process at relatively  high temperatures $T_q\sim 5\;K$.  A
previous study has shown that the normal state  resistivity $\rho_n(T)$ of this crystal (revealed
by the suppression of superconductivity with a large magnetic field) is insulating, similar to that
of $PrBa_{2}Cu_{3}O_{7-\delta}$, so that $\rho_n(T)\rightarrow \infty$ at $T\rightarrow 0$
\cite{Levin}. The reduced dissipation in the normal core of a vortex due to a large normal-state
resistivity increases the mobility of the vortices and facilitates tunneling\cite{G.Blatter:Review}.

The succession of the coupling transition by quantum creep points towards the model of flux flow
shown schematically in the  inset to Fig.~\ref{Arhenn}(b).  Below $T^*$ the vortices begin to form
coherent lines and the dissipation becomes non-ohmic.  The transport current is mostly confined
within a narrow layer below the current contacts. The rest of the sample remains relatively
undisturbed. These two regions of the sample  are uncoupled and, inside each of them, the vortices
are coherent over a distance comparable to the thickness of the respective layer.  In this
scenario, the large resistive anisotropy in the mixed state results from the loss of the phase
coherence only between two macroscopic regions of the sample, not between all microscopic layers
(such as $CuO_2$ bilayers) as in the 2D phase at $T>T^*$.  The fact that at $T<T_q$ we observe a
T-independent $V_p$ and a thermally activated $V_s$ indicates that the thickness of the current-carrying
layer is just a few unit cells (which makes it similar to ultra thin films and multilayers, the only
other systems in which quantum creep was observed in transport measurements).  The small length of
vortices makes tunneling a dominant mechanism even at relatively high temperatures ($T\approx
5\;K$);  the vortices in the lower layer (which is the bulk of the sample) do not tunnel because of
the much longer correlation length and, therefore,  $V_s$ remains thermally activated.

This scenario raises the possibility that the transport current undergoes self-channeling at
$T<T^*$. In the normal and mixed states {\it above} $T^*$, the
$Y_{1-x}Pr_{x}Ba_{2}Cu_{3}O_{7-\delta}$ crystals are not very anisotropic; for  the  crystal
studied (length $L\approx 1\;mm$ and thickness $D\approx 0.015\;mm$), the ratio $V_p/V_s\sim 2$ at
$T>T^*$,  so that the transport current fills the cross-section fairly uniformly. On the other hand,
the transition to quantum creep, which is extremely sensitive to the length of the tunneling
segments, indicates that near  $T_q$, the current-carrying volume collapses into a layer just a few
unit cells thick. A possible reason may be the non-linear as well as nonlocal relationship between 
the current density and electric field in the mixed state below $T^*$. The non-linear resistive
anisotropy may result in instability of the current distribution with respect to formation of a
very thin current-carrying channel. 

In order to confirm that the T-independent $V_p$ [Fig.~\ref{alldata}(a)] is the result of 
quantum flux creep, we performed magnetic relaxation measurements on a similar crystal.
At higher temperatures the relaxation of the magnetic moment proceeds as a
sequence of uncorrelated microscopic steps, each involving thermal activation over a
certain energy barrier $U$. The decay time $\tau_d$ during which the
induced moment loses a {\it substantial fraction} of its initial value
can be defined as:
\be
\label{tau(U,T)}
\tau_{d}=\tau_{esc}\exp\left \{ \frac{U(H,T)}{T}\right \},
\ee
where the Boltzman factor reflects
the degree of availability of energy $U$ required
for an average  elementary step and is essentially independent of
the physics of the relaxation process, while the pre-exponential
factor $\tau_{esc}$ is a measure of how rapidly the 
relaxation would proceed, had it not been limited by
unavailability of thermal energy. We call
$\tau_{esc}$  the {\it escape
time}, to distinguish it from the microscopic attempt time $\tau_a$ which
characterizes the period  of vibration of the vortex 
inside a pinning well. The escape time may depend on the
sample size, magnetic field, and  temperature.
Factorization of $\tau_d$ given by Eq.~(\ref{tau(U,T)}) is meaningful as long as the
Boltzman factor $\exp (U/T)\gg
1$, so that it dominates the $T$ and $H$ dependence
of $\tau_d$.
With $U(H,T)=U_0(H)(1-T/T_{cr})$,
where $T_{cr}$ is the temperature at which the activation energy
vanishes, the decay time becomes:
\be
\label {tau(H,T)}
\tau_{d}=\tau_{esc}\exp\left \{ U_0(H)\left (\frac{1}{T}-\frac{1}{T_{cr}}\right
) \right \}.
\ee
We emphasize that Eq.~(\ref{tau(U,T)}) is more general than any
particular  dynamic model of the relaxation process driven by fluctuations.

Within the decade of time
$10^3-10^4\;s$, the magnetic relaxation curves are well fitted to
$M_{irr}=a-b\ln (t/t_0)$,
where $t_0$ is an arbitrary unit of time. Due to the slowness of the
relaxation, the decay time cannot be
determined directly by monitoring the induced moment until it
loses a substantial fraction of its initial value.  An alternative method, which
we will use, is to estimate the decay time by extrapolating the $M_{irr}(t)$
dependence to $M_{irr}(\tau_d)=0$, which gives:
\be
\label {tau(a,b)}
\tau_d=t_0\exp\{a/b\}.
\ee
With this definition, $\tau_d$ is universal and does not depend on the choice of
$t_0$ so that we can compare the experimental $\tau_d$ with Eq.~\ref{tau(H,T)}.
Figure~\ref{alldata}(b) shows $\tau_d$ calculated according to Eq.~(\ref{tau(a,b)}) and
plotted vs $1/T$ for different values of $H$. 
At higher temperatures the data display an 
Arrhenius dependence with a slope 
$d\ln \tau_d/d(1/T)$ decreasing with increasing $H$.  
The trend is consistent with the
field dependence of the activation energy in transport measurements, Fig.~\ref{alldata}(a).

According to Eq.~\ref{tau(H,T)}, the pre-exponential
factor $\tau_{esc}$ can be determined by the extrapolation of
the Arrhenius  dependence of $\tau_d(T)$ to $T=T_{cr}$.
Indeed, the straight lines extrapolating the activated
dependence converge at $T_{cr}\approx 19\;K\approx T_c$
and $\tau_d=\tau_{esc}\approx 1\;s$.
The value of $\tau_{esc}$ is consistent with the time it takes 
for a vortex to diffuse from the bulk to the outer edge of the
sample: $\tau_{esc}\sim R^2/{\cal{D}}_{v}\sim R^2/\omega_a\ell_a^2$,
where $R$ is the characteristic size of the sample in the direction of
diffusion, 
and ${\cal{D}}_{v}$ is the
diffusion coefficient determined by the attempt frequency $\omega_a$ and the
average hopping distance $\ell_a$. With $\tau_{esc}\sim 1\;s$
and $R^2\sim 10^{-2}-10^{-3}\; cm^2$ (for the crystal we used),
${\cal{D}}_{v}\sim 10^{-2}-10^{-3}\; cm^2/s$. This value of
${\cal{D}}_{v}$ is consistent with an elementary step of the order of the
correlation length $\ell_a\sim 100\;\AA$ and
$\omega_a\sim10^{11}-10^{12}\;s^{-1}$. 

At lower temperatures, the decay time saturates at a roughly $T$- and
$H$-independent level [Fig.~\ref{alldata}(b)]. The values of $T_q(H)$ from
transport[(Fig.~\ref{alldata}(a)] and  magnetic relaxation [Fig.~\ref{alldata}(b)] measurements are
very close in spite of a very large difference in the currents  involved in these
measurements.  The fact that the  transition to a temperature independent
dissipation  takes place in both transport and magnetic relaxation processes,
and at approximately the same temperature in a given field indicates that both
phenomena have the same origin.
The Euclidian action $S_E$ can be estimated as 
$S_E/\hbar=\ln (\tau_d/ \tau_{esc})\approx 25$,
which is comparable, but somewhat  smaller than 
previously reported values for other systems\cite{Hoekstra}.

In summary, we have observed quantum creep in underdoped
$Y_{1-x}Pr_{x}Ba_{2}Cu_{3}O_{7-\delta}$
crystals using both transport and magnetic relaxation measurements. The
transition  to quantum creep is preceded by a coupling transition $T^*$
which leads
to non-ohmic dissipation. The transport data 
indicate that, at $T<T^*$, most of the 
current is confined to a layer just a few unit cells thick
below the current contacts.  This
current-carrying layer is decoupled from the rest of the crystal, where the
vortices are mostly
undisturbed by the current and are coherent over a macroscopically long
distance.
The transition from thermally activated to temperature independent dissipation
takes place only in the current-carrying layer.  The relatively large
probability of tunneling which makes possible the observation of this crossover
at $T \approx 5\;K$  is due to the very short length of the tunneling segments.

This research was supported at KSU by the National Science Foundation under
Grant Nos. DMR-9601839 and DMR-9801990,
and at UCSD by U.S. Department of Energy
under Grant No.
DE-FG03-86ER-45230.

\newpage
%{\bf Figure Captions:}
\end{document}